\documentstyle[aaspp4,psfig]{article}

\def\pp{\par\parshape 2 0truecm 15.5truecm 1truecm 14.5truecm\noindent}
\newcommand{\gtsima}{$\; \buildrel > \over \sim \;$}
\newcommand{\ltsima}{$\; \buildrel < \over \sim \;$}
\newcommand{\simgt}{\lower.5ex\hbox{\gtsima}}
\newcommand{\simlt}{\lower.5ex\hbox{\ltsima}}
\newcommand{\himpc}{{\hbox {$h^{-1}$}{\rm Mpc}} }

\newcommand{\sbkt}[1]{\left(#1\right)}
\newcommand{\abs}[1]{\left|#1\right|}
\newcommand{\ns}{log$N$-log$S$~}
\newcommand{\unit}{erg cm$^{-2}$ s$^{-1}$}
\begin{document}

\title{Constraints on the fluctuation amplitude and density\\
               parameter from X-ray cluster number counts }

\bigskip

\author{Tetsu Kitayama$^{1}$ and Yasushi Suto$^{1,2}$}

\bigskip
\bigskip

\affil{
$^{1}$ Department of Physics, The University of Tokyo, 
Tokyo 113, Japan.\\
$^{2}$ Research Center For the Early Universe (RESCEU), 
School of Science, \\
The University of Tokyo, Tokyo 113, Japan.\\
~\\
e-mail: kitayama@utaphp2.phys.s.u-tokyo.ac.jp, ~
suto@phys.s.u-tokyo.ac.jp}

\bigskip

\centerline{Received 1996 December 20; Accepted 1997 June 27}

\received{1996 December 20}
\accepted{  }

\begin{abstract}
  We find that the observed log$N$ - log$S$ relation of X-ray clusters
  can be reproduced remarkably well with a certain range of values for
  the fluctuation amplitude $\sigma_8$ and the cosmological density
  parameter $\Omega_0$ in cold dark matter (CDM) universes. The
  $1\sigma$ confidence limits on $\sigma_8$ in the CDM models with
  $n=1$ and $h = 0.7$ are expressed as $(0.54 \pm 0.02)
  \Omega_0^{-0.35-0.82\Omega_0+0.55\Omega_0^2}$
  ($\lambda_0=1-\Omega_0$) and $(0.54 \pm 0.02)
  \Omega_0^{-0.28-0.91\Omega_0+0.68\Omega_0^2}$ ( $\lambda_0=0$),
  where $n$ is the primordial spectral index, and $h$ and $\lambda_0$
  are the dimensionless Hubble and cosmological constants. The errors
  quoted above indicate the statistical ones from the observed log$N$ -
  log$S$ only, and the systematic uncertainty from our theoretical
  modelling of X-ray flux in the best-fit value of $\sigma_8$ is about
  15\%.  In the case of $n=1$, we find that the CDM models with
  $(\Omega_0,\lambda_0,h,\sigma_8) \simeq (0.3,0.7,0.7,1)$ and $(0.45,
  0, 0.7, 0.8)$ simultaneously account for the cluster log$N$ -
  log$S$, X-ray temperature functions, and the normalization from the
  {\it COBE} 4 year data. The derived values assume the observations
  are without systematic errors, and we discuss in details other
  theoretical uncertainties which may change the limits on $\Omega_0$
  and $\sigma_8$ from the log$N$ - log$S$ relation.  We have shown the
  power of this new approach which will become a strong tool as the
  observations attain more precision.
\end{abstract}

\keywords{ cosmology: theory -- dark matter -- galaxies: clusters:
  general -- X-rays: galaxies }

\bigskip
\bigskip

\centerline{\sl Accepted for publication in The Astrophysical
  Journal.} 

\newpage

\section{Introduction}

X-ray temperature and luminosity functions (hereafter XTF and XLF) of
galaxy clusters provide important information on cosmology for various
reasons; physics of the X-ray emission from clusters of galaxies is
well understood, and a phenomenological model describing the
temperature and density of intracluster gas, e.g., isothermal
$\beta$-model, is reasonably successful. The dynamical time-scale of
typical clusters is only an order of magnitude smaller than the age of
the universe, but is much shorter than its cooling time-scale (except
at the central core). This implies that such clusters retain the
cosmological conditions at the epoch of their formation without being
affected appreciably by the subsequent physical processes.
Furthermore, one has a theoretical formalism to compute mass functions
of virialized objects fairly reliably (Press \& Schechter 1976,
hereafter PS) which can be applied to predicting the XTF and XLF in a
variety of cosmological models under reasonable assumptions of cluster
evolution.

This methodology is particularly useful in estimating the amplitude of
the density fluctuations. For instance, White, Efstathiou \& Frenk
(1993) found that $\sigma_8$, the rms linear fluctuation in the mass
distribution on a scale $8\himpc$ ($h$ is the Hubble constant $H_0$ in
units of 100 km s$^{-1}$ Mpc$^{-1}$), is approximately given by
$\sigma_8 \sim 0.57\Omega_0^{-0.56}$ in the cold dark matter (CDM)
models with $\lambda_0=1-\Omega_0$, where $\Omega_0$ is the density
parameter and $\lambda_0$ is the dimensionless cosmological constant.
More recently, several authors discussed the constrains on $\Omega_0$
and $\lambda_0$ from the evolution of XTF and XLF (Kitayama \& Suto
1996a,b, hereafter Papers I and II; Viana \& Liddle 1996; Eke, Cole \&
Frenk 1996; Oukbir, Bartlett \& Blanchard 1997).

The most commonly used XTF (Henry \& Arnaud 1991) is, however,
estimated from a small number of clusters $\sim 20$, which ranges less
than one order of magnitude in temperature, and hence the deduced
constraints are statistically limited.  On the contrary, the cluster
number counts, \ns, from recent observations (e.g., Rosati \& Della
Ceca 1997; Ebeling et al. 1997b) are constructed from a sample of
hundreds of clusters and cover almost four orders of magnitude in
flux. While Evrard \& Henry (1991) and Blanchard et al. (1992)
predicted the \ns of X-ray clusters, our present study compares the
latest ROSAT observation with quantitative predictions in very
specific cosmological models, and examines extensively several
systematic uncertainties due to the theoretical modelling.  Our main
finding is that the latest \ns data can be reproduced well in CDM
universes with a set of $(\Omega_0,\lambda_0,\sigma_8)$ which
simultaneously account for the X-ray temperature functions, and the
{\it COBE} 4 year data.

The anisotropies in the microwave background detected by {\it COBE}
offer another independent way of estimating the fluctuation amplitude
(e.g., Bunn \& White 1997).  The resulting estimate of $\sigma_8$ is,
however, very sensitive to the value of the spectral index $n$ of the
primordial fluctuation spectrum as well as $\Omega_0$ and $\lambda_0$,
because the scale probed by {\it COBE} ($\sim 1$Gpc) is about two
orders of magnitude larger than $8\himpc$. On the other hand,
$\sigma_8$ from the cluster abundance is fairly insensitive to the
assumed value of $n$ because clusters are more directly related to the
density fluctuations around $10\himpc$. These two methods are,
therefore, complementary in constraining cosmological models. In this
paper, we adopt $n=1$ for definiteness and derive limits on $\sigma_8
$ and $\Omega_0$ in $\lambda_0=1-\Omega_0$ and $\lambda_0=0$ CDM
universes from the cluster \ns.

\section{Theoretical prediction of the X-ray cluster number counts}

We compute the number of clusters observed per unit solid angle with
flux greater than $S$ by
\begin{equation}
  N(>S) = \int_{0}^{\infty}dz ~d_A^2(z) c \abs{\frac{dt}{dz}}
  \int_{S}^\infty dS_0 ~ (1+z)^3 n_M(M,z)
  \frac{dM}{dT}\frac{dT}{dL_{\rm band}} \frac{dL_{\rm band}}{dS_0},
\label{eq:logns}
\end{equation}
where $c$ is the speed of light, $t$ is the cosmic time, $d_A$ is the
angular diameter distance, $T$ and $L_{\rm band}$ are respectively the
temperature and the band-limited luminosity of clusters, and
$n_M(M,z)dM$ is the comoving number density of virialized clusters of
mass $M \sim M+dM$ at redshift $z$. To be strict, the redshift $z$ at
which one observes a cluster should be conceptually distinguished from
its formation redshift $z_f$.  There exist some formalisms to take
account of the difference explicitly (e.g., Lacey \& Cole 1993,
hereafter LC; Papers I and II). In applying these formalisms, however,
one needs an appropriate theory on the evolution of intracluster gas
in each cluster between $z_f$ and $z$, which is still highly uncertain
and model-dependent at present. In the current analysis, therefore, we
primarily use the standard PS theory to calculate $n_M(M,z)$ assuming
$z_f=z$, and combine it with a phenomenological model of intracluster
gas based upon the observed $L-T$ correlation.  The effect of $z_f
\neq z$ will be also discussed separately on the basis of the LC model
following Paper II.

Given the observed flux $S_0$ and the redshift $z$ of a cluster, we
evaluate its luminosity $L_{\rm band}$, temperature $T$, and mass $M$
in the following manner.  If the observed flux $S_0$ in equation
(\ref{eq:logns}) is given in a band [$E_a$,$E_b$], the source
luminosity $L_{\rm band}$ at $z$ in the corresponding band
[$E_a(1+z)$,$E_b(1+z)$] is written as
\begin{equation}
  L_{\rm band}[E_a(1+z),E_b(1+z)] = 4 \pi d_L^2(z) S_0[E_a,E_b],
\end{equation}
where $d_L$ is the luminosity distance. We then solve the following
equations iteratively to obtain $T$:
\begin{equation}
  L_{\rm bol} = L_{44} \sbkt{\frac{T}{6{\rm keV}}}^{\alpha}
  (1+z)^\zeta ~~ 10^{44} h^{-2}{\rm ~ erg~sec}^{-1},
\label{eq:lt}
\end{equation}
\begin{equation}
  L_{\rm band}[T,E_a(1+z),E_b(1+z)] = L_{\rm bol}(T)\times
  f[T,E_a(1+z),E_b(1+z)],
\label{eq:band}
\end{equation}
where $f[T,E_1,E_2]$ is the band correction factor to translate the
bolometric luminosity $L_{\rm bol}(T)$ into $L_{\rm band}[T,E_1,E_2]$,
and $L_{44}$, $\alpha$ and $\zeta$ are parameters which will be
described shortly.  In computing $f$, we take account of metal line
emissions (Masai 1984) assuming the metallicity of 0.3 times the solar
value, in addition to the thermal bremsstrahlung; the former makes
significant contribution to the soft band luminosity especially at low
temperature and is important for the present study where we use the
ROSAT energy band, $E_a=0.5$keV and $E_b=2.0$keV.  Finally, assuming
that the cluster gas is isothermal, we relate the temperature $T$ to
the mass $M$ by
\begin{eqnarray}
  k_{\rm B} T &=& \gamma {\mu m_p G M \over 3 r_{\rm vir}} \nonumber
  \\ &=& 5.2 ~\gamma \sbkt{{\Delta_{\rm vir} \over 18\pi^2}}^{1/3}
  \sbkt{{M \over 10^{15} M_\odot} }^{2/3} (1+z_f) ~(\Omega_0
  h^2)^{1/3} ~ {\rm keV},
\label{eq:tm}
\end{eqnarray}
where $\gamma$ is a parameter described later, $k_{\rm B}$ is the
Boltzmann constant, $G$ is the gravitational constant, $m_p$ is the
proton mass, and $\mu$ is the mean molecular weight (we adopt
$\mu=0.59$ throughout this paper). The virial radius $r_{\rm
  vir}(M,z_f)$ is computed from $\Delta_{\rm vir}$, the ratio of the
mean cluster density to the mean background density of the universe at
$z_f$. We evaluate the latter using the formulae for the spherical
collapse model presented in Paper II. The above methodology can be
used to predict XTF and XLF as well.  Except in considering the LC
model discussed below, we set $z_f=z$ in the present analysis.

\begin{figure}
\begin{center}
  \leavevmode\psfig{figure=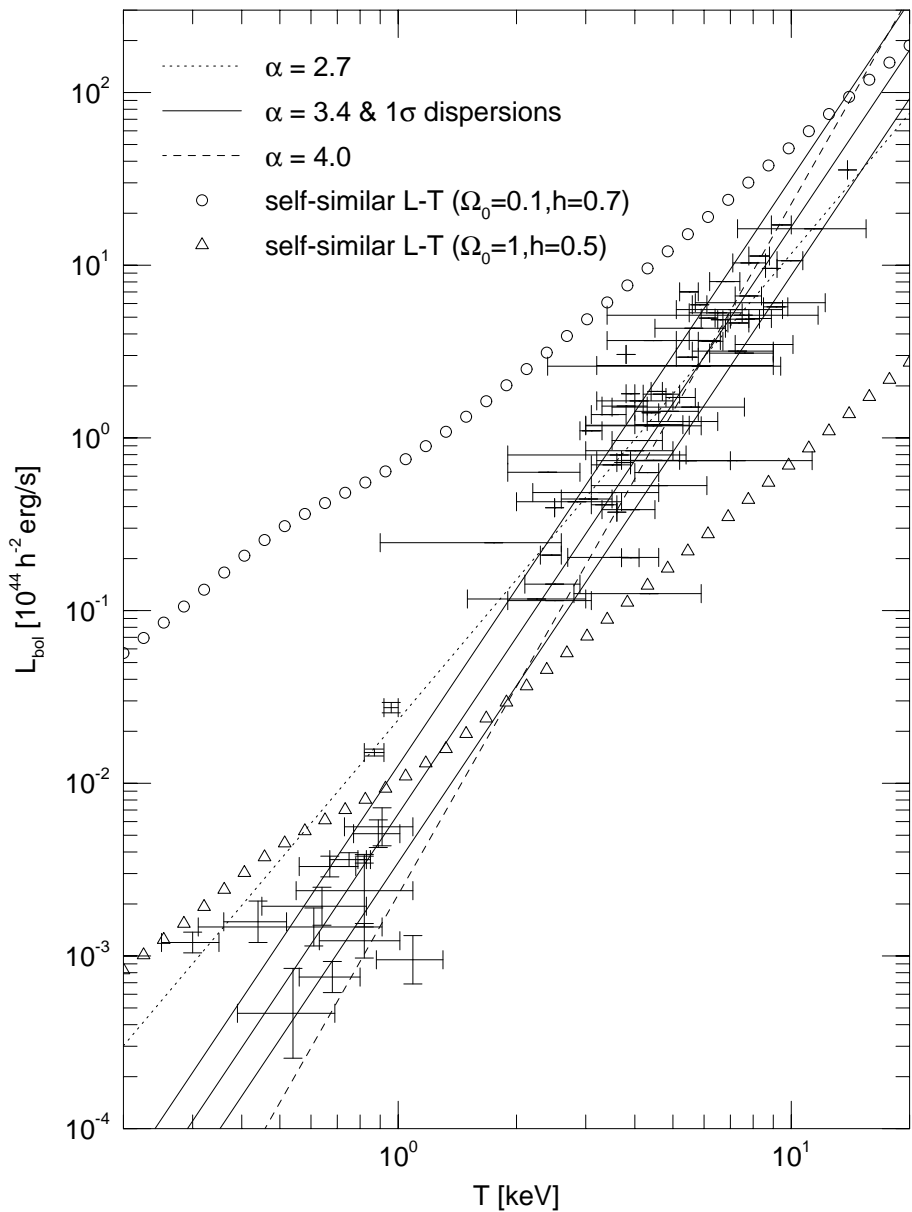,height=16cm}
\end{center}
\caption{The $L-T$ relation of X-ray clusters (at $z=0$).
  The data points at $T\simgt 1.5$keV are from X-ray brightest
  Abell-type clusters (XBACs, Ebeling et al. 1996) while those at
  $T\simlt 1 $keV are from Hickson's compact groups (HCGs, Ponman et
  al.  1996).  Solid lines show our canonical $L-T$ relation
  (\protect\ref{eq:lt}\protect) with $L_{44}=2.9$ and $\alpha=3.4$
  (David et al. 1993) and its $1\sigma$ scatter computed from the data
  points. Dotted line shows the $L-T$ relation with $L_{44}=2.9$ and
  $\alpha=2.7$, and the dashed line with $L_{44}=2.9$ and $\alpha=4$.
  Also plotted are the theoretical $L-T$ relations based on the
  self-similar assumption for $(\Omega_0,\lambda_0,h)=(1,0,0,5)$ (open
  triangles) and $(0.1,0,0,7)$ (open circles). }
\label{fig:lt}
\end{figure}

The above procedure has four parameters; $L_{44}$, $\alpha$ and
$\zeta$ in the $L-T$ relation (\ref{eq:lt}), and $\gamma$ in the $T-M$
relation (\ref{eq:tm}). For $L_{44}$ and $\alpha$, we adopt as our
canonical choice $L_{44}=2.9$ and $\alpha=3.4$ from the observed
present-day $L-T$ relation of David et al. (1993).  We separately
consider the cases of $L_{44}= 1.5$ and $5.5$, and $\alpha=2.7$ and
$4$ in order to take account of the observed scatter to some extent.
Figure \ref{fig:lt} compares our model $L-T$ relation at $z = 0$ with
recent observations; the data at higher temperatures ($T\simgt
1.5$keV) are of the X-ray brightest Abell-type clusters (XBACs;
Ebeling et al. 1996) and the ones at lower temperatures ($T\simlt 1
$keV) are of Hickson's compact groups (HCGs; Ponman et al. 1996). For
both samples, we only plot the clusters with measured X-ray
temperatures (73 clusters from XBACs and 16 from HCGs).  For XBACs, we
adopt the X-ray temperatures from the compilation of David et al.
(1993) and convert the 0.1-2.4 keV band fluxes of Ebeling et al.
(1996) to the bolometric luminosities using the Masai model (1984).

Figure \ref{fig:lt} shows that our model $L-T$ relation (eq.~[3] with
$L_{44}=2.9$ and $\alpha=3.4$) is consistent with the observations of
rich clusters and even small groups over almost two orders of
magnitude in temperature. The best fit value for the slope $\alpha$ of
the $L-T$ relation remains almost unchanged; $\alpha=3.5$ from the
XBACs sample only, and $\alpha=3.3$ from combined samples of XBACs and
HCGs. The range over which we vary $L_{44}$, $1.5 \sim 5.6$,
corresponds to the $\pm 1\sigma$ scatter of the observed data when
$\alpha$ is fixed to 3.4. Figure \ref{fig:lt} also shows that varying
$\alpha$ from $2.7$ to $4.0$ while fixing $L_{44}=2.9$ roughly covers
the scatter in the currently available $L-T$ data for the low
temperature systems.

The parameter $\zeta$ specifies the redshift evolution of the $L-T$
relation. Since recent observations find little evidence for the
evolution in the $L-T$ relation at $z \simlt 0.4$ (e.g., Henry, Jiao
\& Gioia 1994; Mushotzky \& Scarf 1997), we take $\zeta=0$ (no
evolution) as canonical, and also examine the cases of mild evolution
$\zeta= -1$ and $1$ to bracket the possible evolutionary effect.

The value of $\gamma$ in the $T-M$ relation (\ref{eq:tm}) would depend
on the density profile of clusters as well as the ratio of galaxy
kinetic energy to gas thermal energy. In fact, this parametrization
with a single value of $\gamma$, common in the analysis with XTF, may
be too simplified to represent the actual clusters of galaxies, but we
also adopt this in this paper for simplicity. Previous authors mostly
adopt values ranging from $1$ to $1.5$; $\gamma=1$ (Papers I and II),
$\gamma = 1.1$ (Viana \& Liddle 1996), and $\gamma = 1.5$ (Eke et
al. 1996). Recent observations seem to be roughly consistent with this
range, though the scatter is admittedly large (Edge \& Stewart 1991;
Squires et al. 1996; Markevitch et al. 1996). Hereafter, unless
otherwise stated, we adopt $\gamma=1.2$ on the basis of the results of
gas dynamical simulations by White et al. (1993), but again examine
the cases of $\gamma=1$ and $1.5$ so as to see the systematic
uncertainty due to this simplification.

For comparison, we also consider a theoretical $L-T$ relation based on
the self-similar assumption (Kaiser 1986; Paper II). The thermal
bremsstrahlung (free-free) component of the luminosity predicted in
this model is
\begin{equation}
  L_{\rm bol}^{\rm ff} = 1.2 \times 10^{45} \frac{A}{\gamma^{3/2}}
  \sbkt{\frac{\Omega_{\rm B}/\Omega_0}{0.1}}^2 \sbkt{{\Delta_{\rm vir}
      \over 18\pi^2}}^{1/2} \sbkt{\frac{T}{6{\rm keV}}}^2
  (1+z_f)^{3/2} ~(\Omega_0 h^2)^{1/2} ~~ {\rm ~ erg~sec}^{-1},
\label{eq:thlt}
\end{equation}
where $\Omega_{\rm B} =0.0125 h^{-2}$ is the baryon density parameter
(e.g, Walker et al. 1991), and $A$ is a fudge factor of order unity
which depends on the specific density profile of intracluster gas. For
the conventional $\beta$-model profile (eqs.~[3.5]-[3.7] of Paper II),
$A$ is equal to $0.86$ in the case of
$(\Omega_0,\lambda_0,h)=(1,0,0.5)$, and $1.1$ in the case of
$(\Omega_0,\lambda_0,h)=(0.1,0,0.7)$.  In practice, we also take
account of metal line emissions (Masai 1984) in addition to the
free-free component given above. Keeping in mind that the slope of the
self-similar $L-T$ relation is apparently inconsistent with the
observations as summarized in Figure \ref{fig:lt}, we simply intend to
show the results of the simplest theoretical model.  Figure
\ref{fig:lt} clearly exhibits that the amplitude of $L$ in the
self-similar model depends sensitively on the value of $\Omega_{\rm
  B}/\Omega_0$, i.e., the gas mass fraction of the cluster. The
approach based on the observed $L-T$ relation (eq.~[\ref{eq:lt}]), on
the contrary, is entirely independent of it.

\begin{figure}
\begin{center}
  \leavevmode\psfig{figure=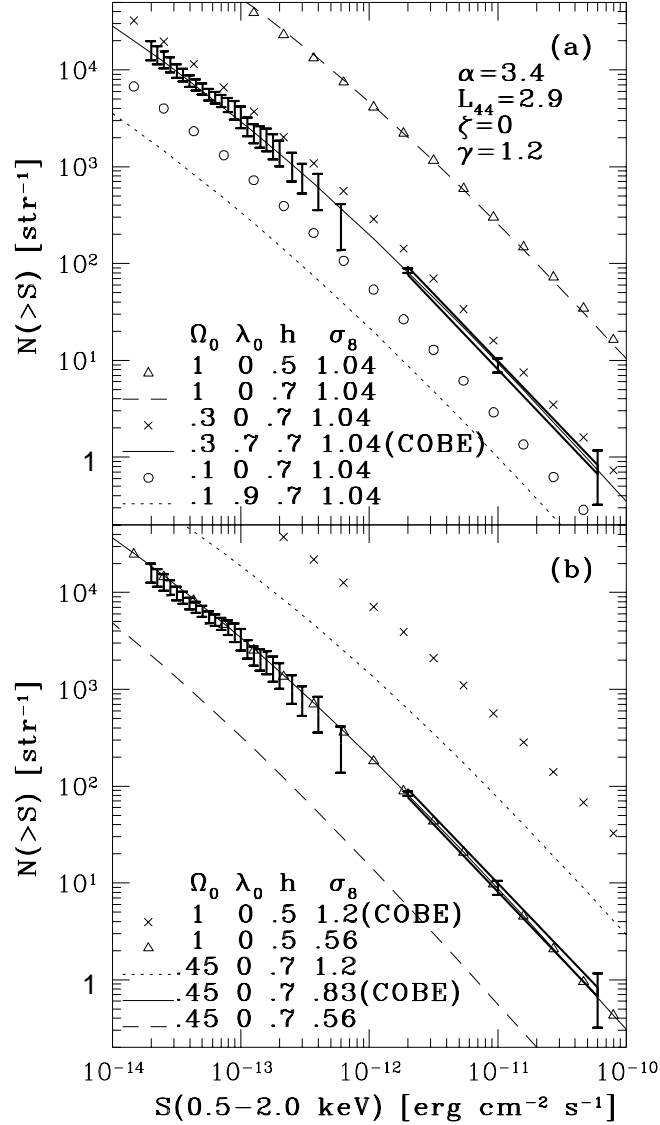,height=16cm}
\end{center}
\caption{ Theoretical predictions for \ns of X-ray
  clusters in CDM models with different cosmological parameters; (a)
  $\sigma_8=1.04$ models with different $\Omega_0$, $\lambda_0$ and
  $h$, (b) $\Omega_0=1$ and $0.45$ models with different
  $\sigma_8$. Denoted by (COBE) are the models normalized according to
  the {\it COBE} 4 year data (Bunn \& White 1997).  Data points with
  error bars at $S\simlt 10^{-12}$ \unit are from the {\it ROSAT} Deep
  Cluster Survey (RDCS, Rosati et al. 1995; Rosati \& Della Ceca
  1997), and the error box at $S\simgt 2 \times 10^{-12}$ represents a
  power-law fitted region from the {\it ROSAT} Brightest Cluster
  Sample (BCS, Ebeling et al. 1997a,b). For the BCS data at $S= 2
  \times 10^{-12}$, $1 \times 10^{-11}$ and $6 \times 10^{-11}$\unit,
  we also plot the corresponding Poisson errors.}
\label{fig:ns1}
\end{figure}

In Figure \ref{fig:ns1}, we plot our predictions of the cluster \ns in
various CDM models. We use the PS mass function in equation
(\ref{eq:logns}) and adopt our canonical set of parameters
($\alpha=3.4$, $L_{44}=2.9$, $\zeta=0$, $\gamma=1.2$) to evaluate the
X-ray flux. We use our fitting formulae (Paper II) for the CDM mass
fluctuation spectrum on the basis of Bardeen et al. (1996) transfer
function. The observed data at fainter fluxes ($S < 10^{-12}$\unit)
are taken from the {\it ROSAT} Deep Cluster Survey (RDCS, Rosati et
al. 1995; Rosati \& Della Ceca 1997) and those at brighter fluxes ($S
> 10^{-12}$\unit) are from the {\it ROSAT} Brightest Cluster Sample
(BCS, Ebeling et al. 1997a,b).  In the analysis below, we use the RDCS
data of Rosati \& Della Ceca (1997) including the systematic errors
according to Rosati (private communication) in addition to the
statistical errors.  The systematics come from the incompleteness in
the optical identification of clusters at faint flux levels and from
uncertainty in the flux determination on the basis of the wavelet
analysis. The former would typically increase the upper error bar by
$+ 15\%$ of $N(>S)$ at a given flux in the range of $2\simlt S \simlt
3 \times 10^{-14}$\unit, while the latter would change $S$ typically
by $+8\%$ which is converted in the error of $N(>S)$.  The error box
for the BCS data is drawn from the best-fit power-law representation
of the data (Ebeling, private communication).  Since this error box
simply represents the fitting errors, we assign the $\pm 1\sigma$
Poisson error (error bars at $S > 10^{-12}$ \unit) estimated from the
number of clusters in the BCS at a given $S$ (the survey area of the
BCS is 4.136 steradian). The Poisson error is used in the statistical
analysis in \S 3.

Figure \ref{fig:ns1}(a) displays the effect of different $\Omega_0$,
$\lambda_0$ and $h$ for $\sigma_8=1.04$ models, while Figure
\ref{fig:ns1}(b) shows that of different $\sigma_8$ in the cases of
$\Omega=1$ and $0.45$.  We find that the {\it COBE} normalized CDM
models with $(\Omega_0,\lambda_0,h,\sigma_8) = (0.3,0.7,0.7,1.04)$ and
(0.45,0,0.7,0.83) reproduce remarkably well the observed \ns over
almost four orders of magnitude in flux provided that we adopt the
canonical set of parameters. In the case of the standard CDM model
with $(\Omega_0,\lambda_0,h) = (1,0,0.5)$, however, the {\it COBE}
normalization ($\sigma_8=1.2$) significantly overproduces the number
of clusters. This discrepancy becomes even worse for $h>0.5$ where the
{\it COBE} normalized $\sigma_8$ becomes larger. Therefore the
standard CDM model is compatible with the cluster number counts only
if $\sigma_8=0.56$, more than a factor of 2 smaller than the COBE
normalization (standard CDM models with $n <1$ can be consistent with
both the COBE and \ns, but we do not explore the possibility in this
paper).  The predicted \ns is sensitive to the values of $\Omega_0$
and $\sigma_8$, but rather insensitive to $\lambda_0$ and $h$.

\begin{figure}
\begin{center}
  \leavevmode\psfig{figure=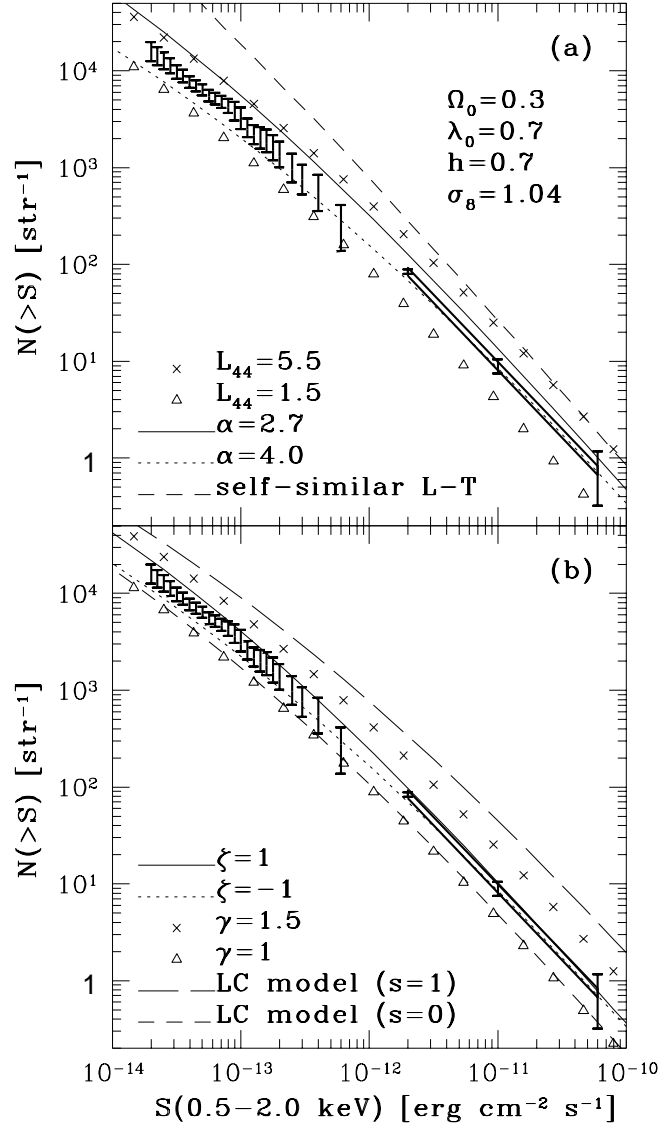,height=16cm}
\end{center}
\caption{ Theoretical predictions for \ns of X-ray
  clusters in a CDM model ($\Omega_0=0.3$, $\lambda_0=0.7$, $h=0.7$,
  $\sigma_8=1.04$); (a) with different $L_{44}$ and $\alpha$ as well
  as the theoretical $L-T$ relation based on the self-similar
  assumption, (b) with different $\zeta$ and $\gamma$ as well as the
  LC model characterised by $T(z) \propto [(1+z_f)/(1+z)]^s$ (see main
  text). }
\label{fig:ns2}
\end{figure}

Figure \ref{fig:ns2} exhibits how the predicted \ns depends on
different choices of parameters to model the $L-T$ and $T-M$
relations. We also plot the results based on the LC model, instead of
the standard PS theory, in evaluating the mass function of virialized
clusters. This model predicts the number of clusters of a given mass
with explicitly taking account of their formation epochs $z_f$ and the
subsequent evolution. We assume that the temperature evolution of
individual clusters is proportional to $[(1+z_f)/(1+z)]^s$ after $z_f$
(see Paper II for details). Figures \ref{fig:ns2} (a) and
\ref{fig:ns2} (b) indicate that varying $\alpha$ or $\zeta$ mainly
changes the slope of the predicted \ns, while different values of
$L_{44}$, $\gamma$ or $s$ affect the amplitude as well and may shift
the \ns predictions by a factor up to $5$.

\section{Constraints on $\Omega_0$ and $\sigma_8$ in CDM models}

\begin{figure}
\begin{center}
   \leavevmode\psfig{figure=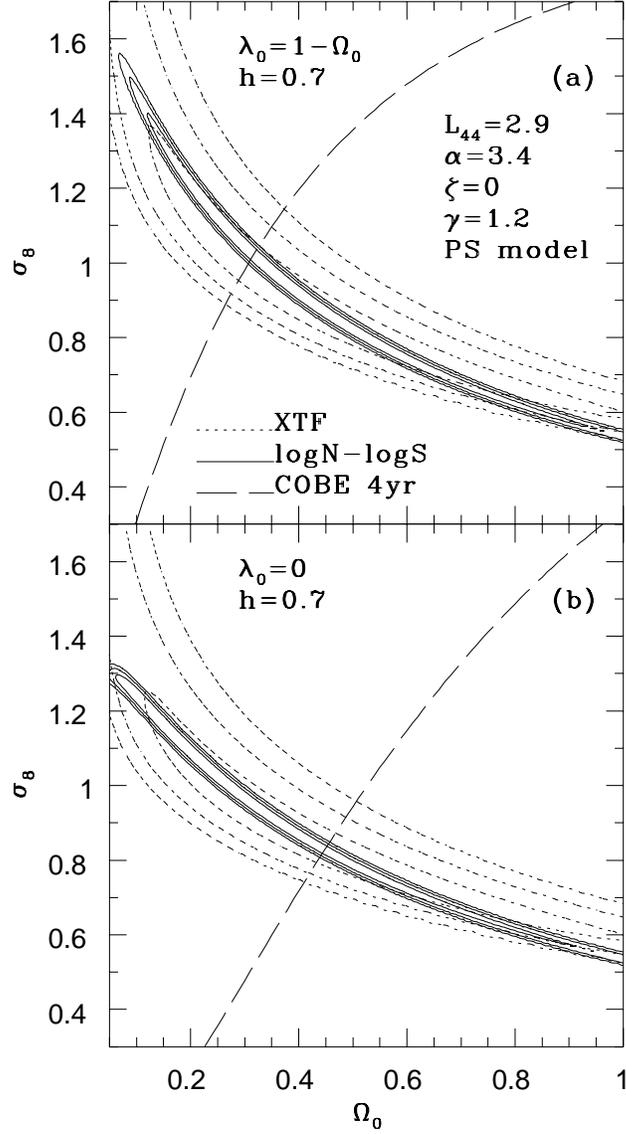,height=16cm}
\end{center}
\caption{Limits on $\Omega_0$ and $\sigma_8$ in CDM models 
  ($n=1$, $h=0.7$) with (a) $\lambda_0=1-\Omega_0$, and (b)
  $\lambda_0=0$. Constraints from cluster \ns (solid) and XTF (dotted)
  are plotted as contours at $1 \sigma$(68\%), $2\sigma$(95\%) and
  $3\sigma$(99.7\%) confidence levels. Dashed lines indicate the {\it
    COBE} 4 year results from Bunn \& White (1997).  }
\label{fig:chi2}
\end{figure}

Figure \ref{fig:chi2} summarizes the constraints on $\sigma_8$ and
$\Omega_0$ from cluster \ns, XTF and {\it COBE} 4 year results (Bunn
\& White 1997) in CDM universes with $h=0.7$ and our standard cluster
model ($\alpha=3.4$, $L_{44}=2.9$, $\zeta=0$, $\gamma=1.2$). We
perform a $\chi^2$ test of the \ns using the six data points; at
$S[\mbox{0.5-2.0 keV}]=4\times 10^{-14}$, $1.2 \times 10^{-13}$ and $3
\times 10^{-13}$\unit from RDCS (Rosati \& Della Ceca 1997), $2 \times
10^{-12}$, $1 \times 10^{-11}$ and $6 \times 10^{-11}$\unit from BCS
(Ebeling et al. 1997b) with appropriate statistical (and systematic)
errors as discussed in the previous section. Strictly speaking, each
data point of the cumulative number counts discussed here is not
independent, but we treat all the above data points as independent.
Since we have selected the six data points where the cluster numbers
are different by a factor of $3 \sim 10$ from their neighboring
points, we expect that this is not a bad approximation. In fact, we
found that the constraint on $\Omega_0$ and $\sigma_8$ plane is
essentially determined by the two data points; at
$2\times10^{-12}$\unit from the BCS survey and at
$4\times10^{-14}$\unit from the RDCS.  We would like to use the six
data points because they would provide some additional information.
Also we repeated the same analysis using twelve data points at
different fluxes and made sure that the resulting constraints are
insensitive to the choice.  For comparison, we also perform a $\chi^2$
test using the XTF data points and associated errors at $T=3$, $4.2$,
and $6.2$keV from Eke et al. (1996) who reanalysed the original data
of Henry \& Arnaud (1991).

Figure \ref{fig:chi2} indicates that constraints from the cluster \ns
are consistent with, but stronger than, those from the XTF, because
the observed \ns has smaller error bars than the XTF and covers wider
dynamic range. Our $1\sigma$ (68\%) confidence limits from cluster \ns
are well fitted by
\begin{eqnarray}
  \sigma_8 = (0.54 \pm 0.02 ) \times \left\{
      \begin{array}{ll}
        \Omega_0^{-0.35-0.82\Omega_0+0.55\Omega_0^2} &
        \mbox{($\lambda_0=1-\Omega_0$)}, \\ 
        \Omega_0^{-0.28-0.91\Omega_0+0.68\Omega_0^2} &
        \mbox{($\lambda_0=0$)}, 
      \end{array}
   \right. 
\label{eq:fit}
\end{eqnarray}
where the quoted errors include only the statistical ones due to the
observed \ns relation (as will be discussed below, the systematic
uncertainty of the above fit due to the theoretical modelling of
cluster luminosity is estimated to be 15\%). The {\it COBE} normalized
$\Omega_0=1$ model is inconsistent with the cluster number counts at
more than $3\sigma$ (99.7\%) confidence.  Observed cluster abundances
and {\it COBE} normalization are simultaneously accounted for by the
CDM model with $(\Omega_0,\lambda_0,\sigma_8) \simeq (0.3,0.7,1)$ and
$(0.45, 0, 0.8)$ in the case of $h=0.7$.

\begin{figure}
\begin{center}
   \leavevmode\psfig{figure=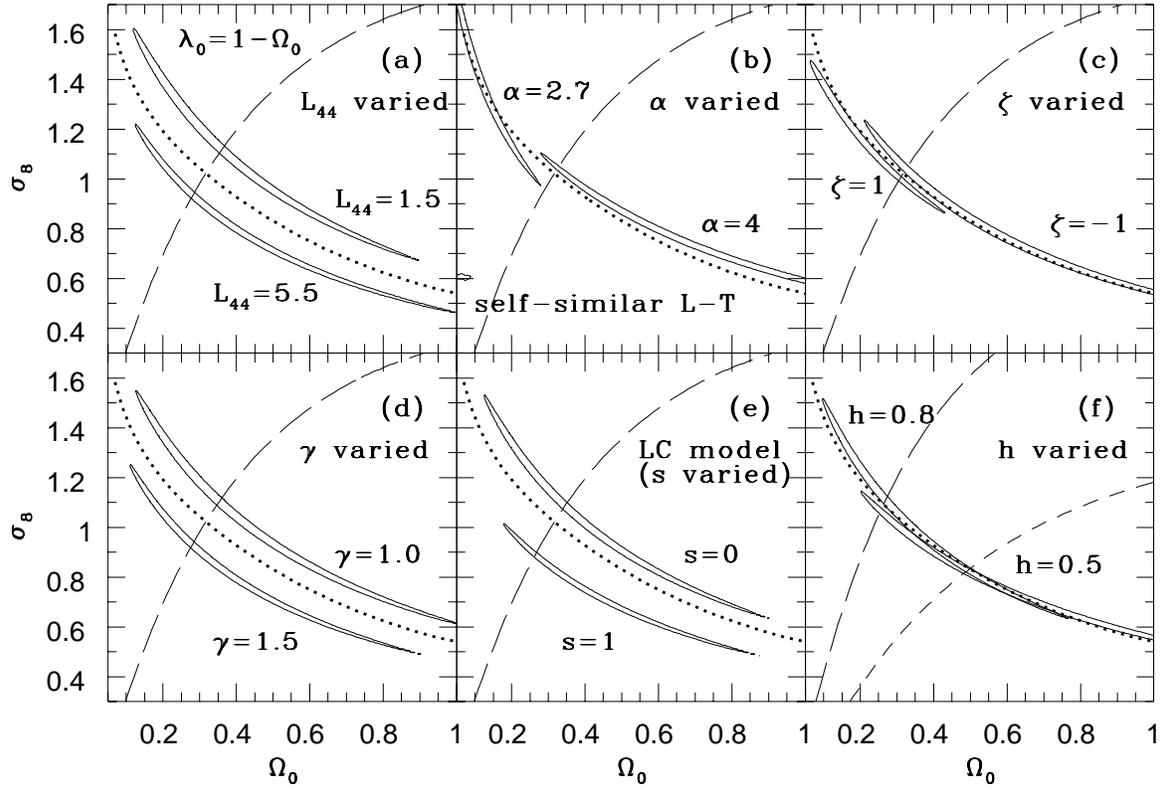,height=16cm}
\end{center}
\caption{Systematic effects on the $\Omega_0 - \sigma_8$ constraints 
  in $\lambda_0=1-\Omega_0$ CDM models. The $1\sigma$(68\%) confidence
  contours from the cluster \ns are plotted for different (a)
  $L_{44}$, (b) $\alpha$, (c) $\zeta$, (d) $\gamma$, (e) $s$, and (f)
  $h$.  Except for the parameters varied in each panel, our canonical
  set of parameters ($L_{44}=2.9$, $\alpha=3.4$, $\zeta=0$,
  $\gamma=1.2$, $h=0.7$) and the PS model are used. Dotted and dashed
  lines represent our best-fit for the canonical parameter set
  (eq.~[\protect\ref{eq:fit}\protect]) and the {\it COBE} 4 year
  results, respectively. }
\label{fig:chi2f}
\end{figure}

\begin{figure}
\begin{center}
  \leavevmode\psfig{figure=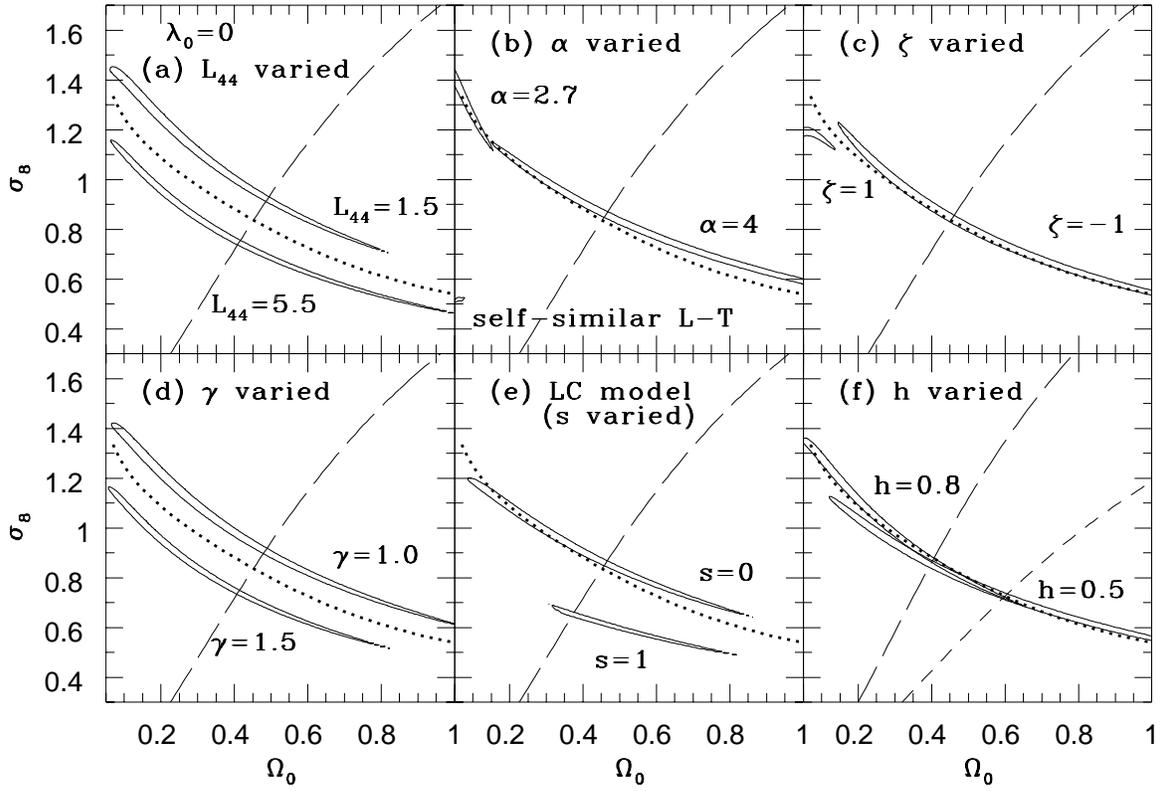,height=16cm}
\end{center}
\caption{Same as Figure \protect\ref{fig:chi2f}\protect, 
  but for $\lambda_0=0$ CDM models.}
\label{fig:chi2o}
\end{figure}

Figures \ref{fig:chi2f} and \ref{fig:chi2o} exhibit the systematic
difference of the above results against our model assumptions in the
$\lambda_0=1-\Omega_0$ and $\lambda=0$ models, respectively. These
figures imply that the dependence of the $\Omega_0-\sigma_8$
constraints on our model parameters are very similar in
$\lambda_0=1-\Omega_0$ and $\lambda=0$ models. Panels (a) indicate
that varying $L_{44}$ from our canonical value $2.9$ to $1.5$ ($5.5$)
will systematically increase (decrease) the best-fit $\sigma_8$ value
for a given $\Omega_0$ by about $15\%$. The range of $L_{44}$
considered here roughly corresponds to the $\pm 1\sigma$ scatter in
the observed $L-T$ relation (see also Fig.\ref{fig:lt}).  Although
this scatter may be partly due to the observational uncertainties in
determining the temperature, we conservatively interpret it as an
intrinsic scatter in the $L-T$ relation which results in the
systematic error for the best-fit $\sigma_8$ value by $15\%$.

Panels (b) to (e) of Figures \ref{fig:chi2f} and \ref{fig:chi2o} show
the systematic effects due to the other model parameters. The best-fit
$\Omega_0-\sigma_8$ relation (eq.~[\ref{eq:fit}]) is shown to be
rather robust against $\alpha$ and $\zeta$ over the ranges considered
here; the changes in these parameters merely move the contours along
the best-fit relation.  This is because $\alpha$ and $\zeta$ mainly
affect the slope of the predicted \ns (Fig.\ref{fig:ns2}) and such
changes are compensated by altering the CDM fluctuation spectrum with
$\Omega_0$ and $\sigma_8$.  On the other hand, allowing for the
changes of $1<\gamma<1.5$ and $0<s<1$, the best-fit $\sigma_8$ value
shifts in a comparable amount to that due to the changes of
$1.5<L_{44}<5.5$.  Note that the changes in $\gamma$ and $s$ change
the XTF and \ns contours in a similar manner, and the resulting
constraints from the \ns and XTF remain consistent with each other.

The ranges of parameters $\alpha$, $\zeta$, $\gamma$ and $s$
considered in Figures \ref{fig:chi2f} and \ref{fig:chi2o} are, unlike
that of $L_{44}$, not directly related to definite statistical
consideration.  Furthermore, it is difficult to judge quantitatively
how their intrinsic uncertainties correlate with one another.  So we
simply illustrate their individual effects in the figures, and quote
only the representative systematic error due to $L_{44}$ for
definiteness and simplicity. As is clear from Figures \ref{fig:chi2f}
and \ref{fig:chi2o}, with the ranges of the parameters considered
here, the error due to $L_{44}$ represents a reasonable estimate for
the total systematic uncertainty in the best-fit $\sigma_8$ value.

Panels (f) of Figures \ref{fig:chi2f} and \ref{fig:chi2o} show that
the cluster number counts is very insensitive to $h$ unlike the {\it
  COBE} normalization. The best-fit cosmological parameters for both
the {\it COBE} data and the cluster abundance are
$(\Omega_0,\lambda_0,\sigma_8) \simeq (0.25,0.75,1.1)$, $(0.4,0,0.9)$
in the case of $h=0.8$, and $(0.5,0.5,0.8)$, $(0.6,0,0.75)$ in the
case of $h=0.5$.

Another interesting application of X-ray cluster number counts can be
found in probing the underlying $L-T$ relation.  Figures
\ref{fig:chi2}, \ref{fig:chi2f} and \ref{fig:chi2o} indicate that the
\ns and XTF contours overlap with each other at the $\pm 1\sigma$
level for our canonical $L-T$ relation with $2.7 < \alpha < 4$ or
$-1<\zeta<1$. On the other hand, \ns and the XLF constraint is in good
agreement with each other only with $L_{44}=2.9$; $L_{44}=1.5$ or
$5.5$ is marginally consistent with the XTF constraint at the
$2\sigma$ level.  Incidentally if we use the theoretical $L-T$
relation briefly described in the previous section, the \ns and XTF
contours do not agree with each other even at the $3\sigma$ level.
These reflect the fact that the predicted \ns is sensitive to the
adopted $L-T$ relation (Fig.~\ref{fig:ns2}).  Thus, with more accurate
determination of the \ns and XTF by the future observations, one will
be able to constrain the $L-T$ relation more tightly.

\section{Conclusions}

We have found that there is a set of theoretical models which
successfully reproduce the observed \ns relation of galaxy clusters
over almost four orders of magnitude in the X-ray flux. This is by no
means a trivial result itself, and more interestingly low density CDM
models with $(\Omega_0,\lambda_0,h,\sigma_8) \simeq (0.3,0.7,0.7,1)$
and $(0.45, 0, 0.7, 0.8)$ in particular simultaneously account for the
cluster \ns, XTF, and the {\it COBE} 4 year results. Constraints on
the density fluctuation spectrum from the abundance of galaxy clusters
are in fact complementary to those from other observations, such as
the cosmic microwave background radiation (Bunn \& White 1997) and the
galaxy correlation functions (Peacock 1997).  Our \ns results confirm
that the {\it COBE} normalized CDM models with $\Omega_0=1$ and
$h\simgt 0.5$ cannot account for the cluster abundances.  The derived
values assume the observations are without systematic errors, and we
discuss in details other theoretical uncertainties which may change
the limits on $\Omega_0$ and $\sigma_8$ from the log$N$-log$S$
relation.  Incidentally these conclusions are also in good agreement
with the recent finding of Shimasaku (1997) on the basis of the X-ray
cluster gas mass function and the big-bang nucleosynthesis
consideration.

Although we have mainly considered CDM models with $n=1$ and $h=0.7$,
our procedure can be easily extended to other cosmological models.
The observed \ns data with better statistical significance can put
more stringent limits on the parameters than the previous estimates
based on the XTF and XLF.  Since the \ns at low fluxes is sensitive to
the underlying $L-T$ relation, one may probe this relation using the
improved data of the \ns and XTF which will become available in the
near future.  In summary, we have shown the power of this new approach
which will become a strong tool as the observations attain more
precision.

\bigskip 
\bigskip 

We deeply thank Piero Rosati, Harald Ebeling, and Patrick Henry for
kindly providing us with the X-ray data prior to their publication. We
also thank Shin Sasaki for stimulating discussion, and an anonymous
referee for useful comments which helped improve the
paper. T.K. acknowledges support from a JSPS (Japan Society of
Promotion of Science) fellowship.  This research was supported in part
by the Grants-in-Aid by the Ministry of Education, Science, Sports and
Culture of Japan (07CE2002) to RESCEU (Research Center for the Early
Universe).

\bigskip
\bigskip

{\it Note Added:} After we submitted this paper, Mathiesen \& Evrard
posted a preprint (astroph/9703176) which also demonstrated the
potential importance of the \ns relation, combined with the
temperature-luminosity relation, for constraining cosmological
parameters. Their results are basically consistent with what we found.

\newpage
\bigskip 
\parskip2pt
\newpage
\centerline{\bf REFERENCES}
\bigskip

\def\apjpap#1;#2;#3;#4; {\pp#1, {#2}, {#3}, #4}
\def\apjbook#1;#2;#3;#4; {\pp#1, {#2} (#3: #4)}
\def\apjppt#1;#2; {\pp#1, #2.}
\def\apjproc#1;#2;#3;#4;#5;#6; {\pp#1, {#2} #3, (#4: #5), #6}

\apjpap Bardeen, J. M., Bond, J. R., Kaiser, N. \& Szalay, A. S. 1986;
  ApJ;304;15;
\apjpap Blanchard, A., Wachter, K., Evrard, A. E., \& Silk, J.
 1992;ApJ;391;1;
\apjpap Bunn, E. F.,\& White, M. 1997;ApJ;480;6;
\apjpap David, L. P., Slyz, A., Jones, C., Forman, W., \& Vrtilek, 
S. D. 1993;ApJ;412;479;  
\apjpap Ebeling, H., Voges, W., B\"{o}hringer, H., Edge, A. C.,
Huchra, J. P., \& Briel, U. G. 1996;MNRAS;281;799; 
\apjpap Ebeling, H., Edge, A. C., Fabian, A. C., Allen, S. W., \&
Crawford C. S. 1997a;ApJ;479;L101;  
\apjppt Ebeling H., et al. 1997b;MNRAS, submitted;  
\apjpap Edge, A. C., \& Stewart, G. C. 1991;MNRAS;252;428;
\apjpap Eke, V. R., Cole, S., \& Frenk, C. S. 1996;MNRAS;282;263;
\apjpap Evrard, A. E., \& Henry, J. P. 1991;ApJ;383;95; 
\apjpap Henry, J. P., \& Arnaud, K. A. 1991;ApJ;372;410; 
\apjpap Henry, J. P., Jiao, L., \& Gioia, I. M. 1994;ApJ;432;49;
\apjpap Kitayama, T., \& Suto, Y. 1996a;MNRAS;280;638 (Paper I);
\apjpap Kitayama, T., \& Suto, Y. 1996b;ApJ;469;480 (Paper II);
\apjpap Lacey, C. G., \& Cole, S. 1993;MNRAS;262;627 (LC);
\apjpap Markevitch, M., Mushotzky, R., Inoue, H., Yamashita, K.,
Furuzawa, A., \& Tawara, Y. 1996;ApJ;456;437;
\apjpap Masai, K. 1984;Ap\&SS;98;367;
\apjppt Mathiesen, B. \& Evrard, A. E. 1997;MNRAS, submitted
(astro-ph/9703176); 
\apjpap Mushotzky, R.F., \& Scharf, C. A. 1997;ApJ;482;L13; 
\apjpap Oukbir, J., Bartlett, J. G., \& Blanchard, A. 1997;A\&A;320;3650;
\apjpap Peacock, J. A. 1997;MNRAS;284;885;
\apjpap Ponman, T. J., Bourner, P. D. J., Ebeling, H., \&
B\"{o}hringer, H. 1996;MNRAS;283;690;
\apjpap Press, W. H., \& Schechter, P. 1974;ApJ;187;425 (PS);
\apjpap Rosati, P., Della Ceca, R., Burg R., Norman, C., \& Giacconi, R. 
1995;ApJ;445;L11;
\apjppt Rosati, P., \& Della Ceca, R. 1997; in preparation;  
\apjppt Shimasaku, K. 1997;submitted to ApJ;
\apjpap Squires, G., Kaiser, N., Babul, A., Fahlman, G., Woods, D.,
Neumann, D.M., B\"{o}hringer, H. 1996;ApJ;461;572;
\apjpap Viana, P. T. P., \& Liddle, A. R. 1996;MNRAS;281;323;
\apjpap Walker T. P., Steigman G., Schramm D. N., Olive K. A., \& 
Kang H. -S. 1991; ApJ; 376; 51;
\apjpap White, S. D. M., Efstathiou, G., \& Frenk, C. S. 1993;MNRAS;262;1023;
\apjpap White, S. D. M., Navarro, J. F., Evrard, A. E., \& Frenk,
C. S. 1993;Nature;366;429;  


\end{document}